\documentstyle[12pt]{article}

\begin{document}

\def\st{1}
\def\dh{2}
\def\bv{3}
\def\ab{4}
\def\bi{5}
\def\bg{6}
\def\at{7}

\def\bk{8}
\def\nt{9}
\def\dl{10}
\def\ew{11}
\def\nn{12}
\def\mn{13}
\def\gs{14}
\def\eg{15}
\def\kn{16}
\def\aw{17}
\def\hn{18}
\def\ot{19}

\begin{titlepage}

\begin{flushright}
{ HU-TFT-93-7 \\
UU-ITP 3/93 \\
hep-th/9301059
}
\end{flushright}

\vskip 0.6truecm

\begin{center}
{\large \bf  ON EXACT EVALUATION OF PATH INTEGRALS \\}
\end{center}

\vskip 1.0cm

\begin{center}
{\bf Antti J. Niemi$^{*}$ } \\
\vskip 0.2cm
{\it Department of Theoretical Physics, P.O. Box 108, S-75108, Uppsala,
Sweden$^{\dag}$ } \\
\vskip 0.3cm
and \\
\vskip 0.3cm
{\it Research Institute for Theoretical Physics, University of
Helsinki \\
Siltavuorenpenger 20 C, SF-00170 Helsinki, Finland  }

\end{center}

\vskip 0.6cm
\begin{center}
{\bf Olav Tirkkonen$^{**}$  } \\
\vskip 0.2cm
{\it Research Institute for Theoretical Physics, University of
Helsinki \\
Siltavuorenpenger 20 C, SF-00170 Helsinki, Finland  }
\end{center}

\vskip 2.3cm

\rm
We develop a general method to evaluate exactly certain phase space
path integrals. Our method is applicable to hamiltonians which are
functions of a classical phase space observable that determines the
action of a circle on the phase space. Our approach is based on the
localization technique, originally introduced to derive the
Duistermaat-Heckman integration formula and its path integral
generalizations. For this, we reformulate the phase space path integral
in an auxiliary field representation that corresponds to a superloop
space with both commuting and anticommuting coordinates. In this
superloop space, the path integral can be interpreted in terms of a model
independent equivariant cohomology, and evaluated exactly in the sense
that it localizes into an integral over the original phase space. The
final result can be related to equivariant characteristic classes.
Curiously, our auxiliary field representation and the corresponding
superloop space equivariant cohomology interpretation of the path
integral essentially coincides with a superloop space formulation of
ordinary Poincare supersymmetric quantum field theories.

\vfill

\begin{flushleft}
\rule{5.1 in}{.007 in}\\
$^{\dag}$ {\small Permanent Address}\\
$^{*}$ {\small E-mail: NIEMI@TETHIS.TEORFYS.UU.SE ~or~
ANIEMI@PHCU.HELSINKI.FI \\}
$^{**}$ {\small E-mail: TIRKKONEN@PHCU.HELSINKI.FI \\ }
\end{flushleft}

\end{titlepage}

\textheight 8.7in

\baselineskip 0.65cm

{\bf 1. Introduction }

\vskip 0.8cm

In the present paper we shall be interested in the exact
evaluation of phase space path integrals. In particular, we
shall identify a general family of hamiltonians for which the
path integral can be evaluated exactly in the
sense, that it {\it localizes} into an integral over
the classical phase space. The final integral may or may not
be evaluated in a closed form. However, since it has a
definite geometric interpretation and is {\it substantially}
simpler than the original path integral, its investigation
using either exact or approximative methods is much
simpler.

Our original motivation comes from
the observations by Semenov-Tyan-Shanskij [\st], and
Duistermaat and Heckman [\dh]. They found, that finite
dimensional phase space integrals of the form
$$
\int \omega^{n} e^{-\beta H}
\eqno (1.1)
$$
where $\omega^{n}$ denotes the Liouville measure, can be {\it
localized} to the critical points of $H$ whenever the
canonical flow of $H$ determines the action
of $U(1) \sim S^{1}$ on the phase space, {\it i.e.}
whenever $H$ can be viewed as a Cartan generator
of some Lie algebra on the phase space. The integration
formula presented in [\st], [\dh] coincides with that
obtained when (1.1) is evaluated by WKB approximation, except
that now the summation is over {\it all} critical points of
the hamiltonian $H$, not just over its local minima as in the
WKB approximation.

Subsequently, it has been observed [\bv-\bi] that the
localization of the integral (1.1)  can be understood in
terms of {\it equivariant cohomology}, hence it is
also intimately related to the concept of {\it equivariant
characteristic classes} [\bg]. The integration formula by
Duistermaat and Heckman has  also been generalized to certain
infinite dimensional cases, and applied in particular to the
evaluation of the Atiyah-Singer index theorem [\at,\bi].

A formal generalization of the Duistermaat-Heckman
integration formula for generic bosonic phase space
path integrals has been presented in [\bk]. The derivation is
based on loop space equivariant cohomology, and the ensuing
integration formula assumes, that the classical
hamiltonian generates the action of $S^{1}$ on the phase
space with isolated critical points. This integration
formula again coincides with that obtained from
the WKB approximation, except that again the summation
extends over all critical trajectories of the classical
action, not just over its local minima. A further
generalization has been presented in [\nt]. This integration
formula  relates the path integral to equivariant
characteristic classes, and the final result can be viewed as
an equivariant version of the Atiyah-Singer index
theorem. In particular, in this generalization the final
result is {\it not } a discrete sum over the critical
trajectories of the action, but an integral over the original
phase space. Consequently it is applicable also in cases,
where the standard WKB approximation does not work, for
example [\dl] if the critical trajectories of the classical
action coalesce at points in the phase space.

The previous integration formulas are all based on
the assumption, that the classical hamiltonian determines
the global action of $S^{1}\sim U(1)$ on the
phase space. A generalization to hamiltonians
that are either bilinear functions of such $U(1)$
generators, or even bilinear functions of arbitrary
generators of some nonabelian Lie algebra, has been
presented in [\ew], and an extension to {\it a priori}
arbitrary functions of such generators has been presented in
[\nn]. The original integrals are now localized to integrals
over some submanifolds of the original phase space, and the
final result can be quite different from the WKB
approximation. As a consequence, these more
general integration formulas provide new
nonperturbative methods for the exact evaluation of a large
class of phase space integrals. They can also be applied to
certain infinite dimensional functional integrals such as
two-dimensional Yang-Mills theory [\ew].

Here we shall generalize
the integration formula presented in [\ew]  to
evaluate exactly certain canonical phase space path
integrals. For this, we shall derive an integration formula
which is applicable whenever the classical hamiltonian is an
{\it a priori} arbitrary function of an observable that
generates the action of $U(1)\sim S^{1}$ on the phase space.
The final result has a definite interpretation in terms of
equivariant characteristic classes. It is an integral over
the classical phase space of the theory, and reduces to the
simple form presented in [\nt] if specified to a hamiltonian
that is a generator of $S^{1}\sim U(1)$.

In the present case, the integration
formula considered in [\ew] can not be directly
applied:  Due to the kinetic term that appears in the
classical action in addition of the hamiltonian, the
integrand of canonical path integral can not be represented
in the simple functional form discussed in [\ew]. In a sense,
our integration formula  can then be viewed as a loop space
generalization of the integration formulas discussed there.
Its derivation is based on loop space equivariant cohomology,
in an auxiliary field representation of the original phase
space path integral with both commuting and anticommuting
coordinates {\it i.e.} a superloop space. Curiously, we find
that the pertinent superloop space essentially coincides with
the superloop space introduced in [\mn], to relate generic
supersymmetric theories to loop space equivariant cohomology.
Furthermore, we find that in analogy with the general
formalism developed in [\mn], the pertinent equivariant
cohomology that we shall use here to evaluate the path
integral is again that of model independent $S^{1}$ action in
the superloop space.

In Section 2. we shall present a review of
symplectic geometry, and its appropriate
generalization to loop space. We also discuss some general
aspects of $S^{1}$ action in the loop space. In Section 3. we
evaluate the path integral for a hamiltonian that generates
a model dependent action of $S^{1}$ in the classical phase
space. Our integration formula localizes the corresponding
path integral into an integral over the phase space
in a manner that has a very definite interpretation in terms
of equivariant characteristic classes.  In
Section 4. we explain, how this result can be related to
model independent loop space $S^{1}$ equivariant cohomology,
defined in an auxiliary superloop space with both commuting
and anticommuting coordinates. In Section 5. we show, that
this model independent loop space $S^{1}$ equivariant
cohomology and the corresponding superloop space
representation of the original path integral can be
generalized to derive an integration formula for path
integrals with a hamiltonian which is quadratic
in a  generator of  $U(1) \sim S^{1}$. In Section 6. we
generalize this integration formula to a hamiltonian which is
an {\it a priori} arbitrary function of such a $U(1)\sim
S^{1}$ generator, and in Section 7. we  verify that in
certain simple cases our integration formula indeed yields
correct results.

\vfill\eject

{\bf 2. Loop Space And Circle Action}

\vskip 0.8cm

In the following we shall be interested in the evaluation of

canonical phase space path integrals

$$
Z ~=~ \int [dz^{a}] \prod\limits_{t}\sqrt{ det||
\omega_{ab} || } exp \{ i \int\limits_{0}^{T} \vartheta_{a} {\dot
z}^{a} - H(z) \} \eqno (2.1)
$$
We shall argue, that if the hamiltonian $H(z)$ satisfies a certain
condition which we shall specify in the following, the path
integral (2.1) can be evaluated exactly in the sense that it
reduces to an ordinary integral over the classical phase space
$\Gamma$.

The appropriate interpretation of (2.1) is in terms of
symplectic geometry [\gs] in a canonical loop space $L\Gamma$ over
the classical phase space $\Gamma$. The symplectic geometry of
$L\Gamma$ is constructed from the symplectic
geometry of $\Gamma$, and for this we
consider $\Gamma$ in a generic coordinate system $z^{a}$
($a=1,...,2n = {\rm dim}(\Gamma)$). In these coordinates, the
fundamental Poisson bracket is

$$
\{ z^{a} , z^{b} \} ~=~ \omega^{ab}(z)
\eqno (2.2)
$$
and the inverse matrix $\omega_{ab}$ that appears in (2.1),
$$
\omega^{ac}\omega_{cb} ~=~ \delta_{b}^{a}
\eqno (2.3)
$$
determines components of the symplectic two-form on the phase
space $\Gamma$,
$$
\omega ~=~ \frac{1}{2}\omega_{ab} dz^{a} \wedge
dz^{b}  \eqno (2.4)
$$
This symplectic two-form is closed,
$$
d\omega ~=~ 0 \eqno (2.5)
$$
or in components,
$$
\partial_{a} \omega_{bc} + \partial_{b}\omega_{ca} +
\partial_{c}\omega_{ab} ~=~ 0
\eqno (2.6)
$$
which is equivalent to the Jacobi identity for the Poisson
bracket (2.2).

{}From (2.5) we conclude, that we can represent
$\omega$ locally as an exterior derivative of a one-form. The
functions $\vartheta_{a}(z)$ that appear in (2.1) are components
of this symplectic one-form,
$$
\omega ~=~ d\vartheta ~=~ \partial_{a}
\vartheta_{b} dz^{a} \wedge dz^{b}
\eqno (2.7)
$$
and smooth, real valued functions $\psi$ on
$\Gamma$ define diffeomorphisms that leave $\omega$
invariant: If we introduce a change of variables
$z^{a} \to \tilde z^{a}$ such that
$$
\vartheta_{a} dz^{a} ~=~ \vartheta ~ {\buildrel {\psi} \over
{\longrightarrow} } ~ \vartheta + d\psi ~=~(\vartheta_{a} +
\partial_{a}\psi)dz^{a} ~=~ \tilde \vartheta_{a} d{\tilde z}^{a}
\eqno (2.8)
$$
we conclude from $d^{2}=0$ that $\omega$ remains intact,
$$
\omega ~ {\buildrel {\psi} \over {\longrightarrow} } ~ \tilde
\omega ~\equiv ~ \omega
\eqno (2.9)
$$
The change of variables (2.8) determines a canonical
transformation on $\Gamma$, and $\psi$ is the generating
function of this transformation. Indeed, Darboux's theorem states
that locally in a neighborhood on $\Gamma$ we can always introduce
a change of variables $z^{a} \to p_{a}, q^{a}$ such that $\omega$
becomes
$$
\omega ~=~ dp_{a}\wedge dq^{a}
\eqno (2.10)
$$
where $p_{a}$ and  $q^{a}$ are canonical momentum
and position variables on $\Gamma$. In these variables the
symplectic one-form becomes
$$
\vartheta ~=~ p_{a} dq^{a}
\eqno (2.11)
$$
and (2.8) becomes
$$
p_{a} dq^{a} = \vartheta ~ {\buildrel {\psi}
\over {\longrightarrow} } ~
\vartheta + d\psi ~=~ \tilde \vartheta ~=~ P_{a} dQ^{a}
\eqno (2.12)
$$
Consequently
$$
p_{a} d q^{a} - P_{a} dQ^{a} ~=~ d\psi
\eqno (2.13)
$$
where both $p_{a}, q^{a}$ and $P_{a}, Q^{a}$ are
canonical momentum and coordinate variables on
$\Gamma$. This is the standard form of a canonical transformation
determined by the generating functional $\psi$.

The exterior products of $\omega$ determine closed $2k$-forms on
$\Gamma$. The $2n$-form (where $ {\rm dim}(\Gamma) = 2n$)
$$
\omega^{n} ~=~ \omega \wedge ... \wedge \omega ~~~~~(n ~ {\rm times})
\eqno (2.14)
$$
defines a natural volume element on $\Gamma$ which is invariant
under canonical transformations (2.8). This is the Liouville
measure, and in local Darboux coordinates (2.10) it becomes the
familiar
$$
\left\{ \frac{1}{n !}(-)^{n(n-1)/2}\right\} \cdot \omega^{n} ~=~
dp_{1} \wedge ...  \wedge dp_{n} \wedge
dq^{1} \wedge ... \wedge dq^{n}
\eqno (2.15)
$$

Smooth, real-valued functions $F$ on $\Gamma$ are called classical
observables. The symplectic two-form associates to the
exterior derivative $dF$ of a classical observable $F$ a
hamiltonian vector field ${\cal X}_{F}$ by
$$
\omega ( {\cal X}_{F} , \circ ) ~+~ dF ~=~ 0
\eqno (2.16)
$$
or in component form
$$
{\cal X}_{F}^{a} ~=~ \omega^{ab}\partial_{b} F
\eqno (2.17)
$$
The Poisson bracket of two classical observables $F$ and $G$ can
be expressed in terms of the corresponding vector fields
$$
\{ F , G \} ~=~ \omega^{ab} \partial_{a}F  \partial_{b}G
{}~=~ {\cal X}_{F}^{a}\partial_{a}G
{}~=~ \omega_{ab} {\cal X}_{F}^{a} {\cal X}_{G}^{b}
{}~=~ \omega({\cal X}_{F} , {\cal X}_{G} )
\eqno(2.18)
$$
This determines the
internal multiplication $i_{F}$ of the one-form $dG$ by the vector
field ${\cal X}_{F}$,
$$
i_{F} dG ~=~ {\cal X}_{F}^{a} \partial_{a}G
\eqno (2.19)
$$
More generally, internal multiplication $i_{F}$ by a vector
field ${\cal X}_{F}$ is a nilpotent operation which is defined on
the exterior algebra $\Lambda$ of the phase space $\Gamma$,
$$
i_{F}^{2} ~=~0
\eqno (2.20)
$$
that maps the space $\Lambda_{k}$ of $k$-forms to the space
$\Lambda_{k-1}$ of $k-1$-forms. Using $d$ and $i_{F}$ we
introduce the equivariant exterior derivative
$$
d_{F} ~=~ d + i_{F}
\eqno (2.21)
$$
defined on the exterior algebra $\Lambda$ of $\Gamma$.
Since $d$ maps the subspace $\Lambda_{k}$ of
$k$-forms onto the subspace $\Lambda_{k+1}$ of $k+1$-forms, (2.21)
does not preserve the form degree but maps an even form onto an
odd form, and vice versa. Hence it can be viewed as a
supersymmetry operator. The corresponding supersymmetry
algebra closes to the Lie derivative ${\cal L}_{F}$ along the
hamiltonian vector field ${\cal X}_{F}$,
$$
d_{F}^{2} ~=~ d i_{F} + i_{F} d
{}~=~ {\cal L}_{F}
\eqno (2.22)
$$
and the Poisson bracket (2.18) coincides with the Lie
derivative of $G$ along the hamiltonian vector  field ${\cal
X}_{F}$,
$$
\{ F , G \} ~=~
\omega^{ab}\partial_{a}F \partial_{b}G ~=~ {\cal X}_{F}^{a}
\partial_{a} G ~=~ {\cal L}_{F} G
\eqno (2.23)
$$

The linear space of $\xi \in \Lambda$ which is annihilated by the
Lie-derivative (2.22),
$$
{\cal L}_{F} \xi ~=~ 0
\eqno (2.24)
$$
determines an invariant subspace $\Lambda_{inv}$ of the
exterior algebra $\Lambda$ which is mapped onto itself by the
equivariant exterior derivative $d_{F}$. The restriction of
$d_{F}$ on this subspace is nilpotent, $d_{F}^{2} =
0$, hence $d_{F}$ determines a conventional
exterior derivative in this subspace. The equivariant
cohomology $H^{\star}_{F}(\Lambda)$ that $d_{F}$ determines in
the exterior algebra of $\Gamma$ can be identified with
its ordinary cohomology $H^{\star}(\Lambda_{inv})$ in this
subspace.

In the following we shall interpret (2.1) in terms of symplectic
geometry in the loop space $L\Gamma$. This loop space is
parametrized by the time  evolution\footnotemark\footnotetext{
Notice that in a field theory application the space coordinates
$\vec{\bf x}$ are contained in the degrees of freedom
characterized by the index $a$ in the usual fashion.}
$z^{a} \to z^{a}(t)$ with periodic boundary conditions $z^{a}(0)
= z^{a}(T)$.  The exterior derivative in $L\Gamma$ is obtained by
lifting the exterior derivative of the phase space $\Gamma$,
$$
d ~=~ \int\limits_{0}^{T} dt ~{dz}^{a}(t) {\delta \over \delta
z^{a} (t) } ~\equiv~ {dz}^{a} {\delta \over \delta z^{a}}
\eqno (2.25)
$$

where $dz^{a}(t)$ denotes a basis of
loop space  one-forms\footnotemark\footnotetext{ In the following
we usually do not write explicitly integration over
the loop space parameter $t$. It will always be clear
from the context if an integral over $t$ is
understood.}, obtained by lifting a basis of one-forms in the
phase space $\Gamma$ to the loop space.

The loop space symplectic geometry is determined by a
loop space symplectic two-form
$$
\Omega ~=~ \int dt dt' ~ \frac{1}{2} \Omega_{ab}(t,t')
{dz}^{a}(t) \wedge  {dz}^{b}(t')
\eqno (2.26)
$$
It is a closed two-form in the loop space,
$$
d \Omega ~=~ 0
\eqno (2.27)
$$
or in local coordinates $z^{a}(t)$,
$$
{\delta \over \delta z^{a}}\Omega_{bc} + {\delta \over \delta
z^{b}} \Omega_{ca}+ {\delta \over \delta z^{c}} \Omega_{ab} = 0
\eqno (2.28)
$$
Hence we can locally represent $\Omega$ as an
exterior derivative of a loop space one-form,
$$
\Omega ~=~ d\Theta
\eqno (2.29)
$$
where
$$
\Theta ~=~ \int dt ~\Theta_{a}(t) {dz}^{a}(t)
\eqno (2.30)
$$
We shall assume that (2.26) is nondegenerate, {\it i.e.}
the matrix $\Omega_{ab}(t,t')$ can be inverted in the loop space.
Examples of such nondegenerate two-forms are obtained
by lifting symplectic two-forms $\omega_{ab}(z)$ from
the original phase space $\Gamma$ to the loop space,
$$
\Omega_{ab}(t,t') ~=~ \omega_{ab}[z(t)] \delta(t-t')
\eqno (2.31)
$$
Similarly other quantities can be lifted from the original phase
space to the loop space.

In particular, we define loop space canonical
transformations as loop space changes of variables, that leave
$\Omega$ invariant. These transformations are of the form
$$
\Theta   ~ {\buildrel {\Psi}
\over {\longrightarrow} } ~\tilde
\Theta ~=~ \Theta + d \Psi
\eqno (2.32)
$$
with $\Psi [z(t)]$ the
generating functional of the canonical transformation.

The exterior products of $\Omega$ determine canonically invariant
closed forms on $L\Gamma$, and the top form
yields a natural volume element, the loop
space Liouville measure. We are particularly interested in
corresponding integrals which are of the form
$$
Z~=~ \int [dz^{a}] \sqrt{ det||\Omega_{ab}|| } exp \{ i
S_{B}\}
\eqno (2.33)
$$
with $S_{B}(z)$ a loop space observable {\it
i.e.} a functional on $L\Gamma$. If we specify (2.31) and identify
$S_{B}$ with the action in (2.1), we can interpret (2.1) as an
example of such a loop space integral.

We exponentiate the determinant
in (2.33) using anticommuting variables $c^{a}(t)$,
$$
Z~=~ \int [dz^{a}] [dc^{a}] exp\{ i S_{B} +
ic^{a}\Omega_{ab}c^{b} \} ~=~ \int [dz^{a}] [dc^{a}] exp\{ i S_{B}
+ iS_{F} \}
\eqno (2.34)
$$
This integral is invariant under a loop space supersymmetry
transformation: If ${\cal X}_{S}^{a}$ is the
loop space hamiltonian vector field determined by the
functional $S_{B}$,
$$
{\delta S_{B} \over \delta z^{a}} ~=~\Omega_{ab}{\cal X}_{S}^{b} \eqno
(2.35)
$$
the supersymmetry is
$$
d_{S} z^{a} ~=~ c^{a} \eqno (2.36.a)
$$
$$
d_{S} c^{a} ~=~ - {\cal X}_{S}^{a} \eqno (2.36.b)
$$
 This supersymmetry can be related to loop space equivariant cohomology.
For this we identify $c^{a}(t)$ as the loop space one-forms $
{dz}^{a}(t) \sim c^{a}(t)$, and introduce the loop space equivariant
exterior derivative
$$
d_{S} ~=~d + i_{S}
\eqno (2.37)
$$
Here $i_{S}$ denotes contraction along the
hamiltonian vector field ${\cal X}_{S}^{a}$,
$$
i_{S} ~=~ {\cal X}_{S}^{a} i_{a}
\eqno (2.38)
$$
and $i_{a}(t)$ is a basis of loop space contractions which is
dual to $c^{a}(t)$,
$$
i_{a}(t) c^{b}(t') ~=~ \delta_{a}^{b}(t-t')
\eqno (2.39)
$$

Again, (2.37) fails to be nilpotent and its
square determines the loop space Lie-derivative along
${\cal X}_{S}^{a}$,
$$
d_{S}^{2} ~=~ di_{S} + i_{S}d ~=~{\cal L}_{S}
\eqno (2.40)
$$

The action in (2.34) is a linear combination of a loop space
zero-form ($S_{B}$) and a two-form ($S_{F}$). The
supersymmetry (2.36) means that it is equivariantly closed in the
loop space,
$$
d_{S} (S_{B} +  S_{F}) ~=~ 0
\eqno (2.41)
$$
Hence the action can be {\it locally} represented as an
equivariant exterior derivative of a one-form $\hat\Theta$,
$$
S_{B} + S_{F} ~=~ (d+i_{S}) \hat\Theta ~=~ \hat\Theta_{a} {\cal
X}_{S}^{a} + c^{a}\Omega_{ab}c^{b}
\eqno (2.42)
$$
and the supersymmetry (2.41) implies that
$$
d_{S}^{2} \hat\Theta ~=~ (d i_{S} + i_{S} d) \hat\Theta ~=~
{\cal L}_{S} \hat\Theta ~=~ 0
\eqno (2.43)
$$
so that $\hat\Theta$ is in the subspace where
$d_{S}$ is nilpotent. If $\Phi_{S}$ is some globally
defined loop space zero-form such that
$$
{\cal L}_{S} (d\Phi_{S}) ~=~ 0
\eqno (2.44)
$$

we conclude that $\hat\Theta$ is not unique but the action (2.42)
is invariant under the loop space canonical transformation
$$
\hat\Theta ~\to~ \hat\Theta + d \Phi_{S}
\eqno (2.45)
$$
Consequently (2.34) has a very definite interpretation in terms
of the loop space cohomology which is determined by $d_{S}$ in the
subspace ${\cal L}_{S} = 0$, where $d_{S}$ is nilpotent.

The supersymmetry (2.41) can be used to derive a loop space
generalization of the Duistermaat-Heckman integration formula
[\bk]
$$
Z ~=~ \int [dz^{a}]\sqrt{Det || \Omega|| } e^{iS_{B}} ~=~
\sum\limits_{\delta S_{B}=0}{  \sqrt{ Det || \Omega || } \over
\sqrt { Det ||\delta^{2} S_{B} ||} } e^{i S_{B}}
\eqno (2.46)
$$
Here the sum on the {\it r.h.s.} is over all critical points of
the action $S_{B}$, {\it i.e.} over the zeroes of the hamiltonian
vector field ${\cal X}_{S}^{a}$. The derivation of
(2.46) assumes [\bk], that the loop space admits a Riemannian
structure with a {\it globally defined} loop space metric tensor
$G_{ab}(z; t,t')$ which is Lie-derived by the vector field ${\cal
X}_{S}^{a}$,
$$
{\cal L}_{S} G ~=~ 0
\eqno (2.47)
$$
or in component form,
$$
\partial_{a} {\cal X}_{S}^{b} G_{bc} + \partial_{c}
{\cal X}_{S}^{b} G_{ba} + {\cal X}_{S}^{b}\partial_{b}G_{ac} ~=~0
\eqno (2.48)
$$
For a {\it compact} phase space the corresponding condition
would mean, that the canonical flow generated
by the hamiltonian vector field ${\cal X}_{S}^{a}$ corresponds
to the global action of a circle $S^{1}\sim U(1)$ on the phase
space $\Gamma$. We shall assume that this is the relevant case also
in the loop space.  We parametrize the $S^{1}$ flow that ${\cal
X}_{S}$ generates by a continuous parameter $\tau$ {\it i.e.}
$z^{a} \to z^{a}(\tau)$ with $z^{a}(1) = z^{a}(0)$. Thus
$$
{\cal X}_{S}^{a}(z[t]) ~=~
{\partial z^{a}(t;\tau) \over \partial \tau
}_{|\tau =0}    \eqno (2.49)
$$
and if we also assume that
we have selected the coordinates $z^{a}(t)$
so that the flow parameter $\tau$ shifts the loop (time) parameter
$t\to t+\tau$ we get
$$
{\cal X}_{S}^{a}(z[t]) ~=~ {\partial
z^{a}(t;\tau) \over \partial \tau }_{|_{\tau=0}} ~=~ {d z^{a}
(t) \over dt} ~\equiv~ {\dot z}^{a} \eqno (2.50)
$$
The relation (2.35) then simplifies to
$$
{\delta S_{B} \over \delta z^{a} }
{}~=~ \Omega_{ab}(z) {\dot z}^{b}
\eqno (2.51)
$$
and the supersymmetry transformation (2.36) becomes
$$
d_{\dot z} z^{a} ~=~ c^{a}
\eqno (2.52.a)
$$
$$
d_{\dot z} c^{a} ~=~\dot z^{a}
\eqno (2.52.b)
$$
Here
$$
d_{S} ~\to~ d_{\dot z} ~=~ d + i_{\dot z}
\eqno (2.53)
$$

is the equivariant exterior derivative along the $S^{1}$-vector
field ${\cal X}_{S}^{a} \to \dot z^{a}$, and the
corresponding Lie derivative is simply
$$
{\cal L}_{\dot z} ~=~ di_{\dot z} + i_{\dot z}d ~\sim~
\frac{d}{dt}
\eqno (2.54)
$$
In particular, we conclude that (locally) the action (2.42) is now
of the functional form
$$
S_{B} + S_{F} ~=~ \int \theta_{a} {\dot z}^{a} + c^{a}\Omega_{ab}
c^{b} ~=~ (d+i_{\dot z})\hat\Theta ~=~ d_{\dot z}\hat\Theta
\eqno (2.55)
$$
and as a consequence of (2.44), (2.54) it is invariant under a {\it
generic} loop space canonical transformation
$$
\hat\Theta ~{\buildrel {\Psi} \over {\longrightarrow} }~
\hat\Theta + d\Psi
\eqno (2.56)
$$
where $\Psi$ is now an {\it arbitrary} globally defined and
single-valued functional on $L\Gamma$. The
corresponding path integral (2.34) is also (at least formally)
invariant under these canonical transformations.

In the following sections we shall argue, that {\it in a proper
auxiliary field formalism} a general class of path integrals can be
described by an action which is of the functional form (2.55).  In the
auxiliary field representation that we shall introduce, these path
integrals then essentially coincide with the formalism discussed here.

\vfill\eject

{\bf 3. Hamiltonians That Generate Circle Action}

\vskip 0.8cm

We shall now evaluate the path integral (2.1) for a
hamiltonian $H$ that generates the global action of
$S^{1}\sim U(1)$ on the classical phase space $\Gamma$. We shall
first evaluate this path integral by interpreting it in

the space of loops that are defined in the original phase space
$\Gamma$. The relevant loop space equivariant exterior
derivative will have the functional form
$$
d + i_{\dot z} + i_{H}
\eqno (3.1)
$$
Notice that this does {\it not} correspond to the action of
$S^{1}$ in the loop space $L\Gamma$ as described in section 2. In
order to formulate the path integral in terms of such a
model independent $S^{1}$-formalism we need to introduce an
appropriate auxiliary field representation. We shall explain this
auxiliary field formalism in the subsequent sections. The present
evaluation of (2.1) will then be quite valuable in guiding us to
correctly evaluate the auxiliary field representation of the path
integral.

As in (2.34), we introduce anticommuting variables $c^{a}$ and
write (2.1) as
$$
Z ~=~ \int [dz^{a}] [dc^{a}] exp \{ i
\int\limits_{0}^{T} \vartheta_{a} \dot z^{a}
- H + \frac{1}{2}c^{a}\omega_{ab}c^{b} \}
\eqno (3.2)
$$
The loop space hamiltonian vector field that
corresponds to the bosonic part of the action is
$$
{\cal X}_{S}^{a} ~=~ \dot z^{a} - \omega^{ab}
\partial_{b} H \eqno (3.3)
$$
and we identify $c^{a}(t)$ as a basis of
one-forms on this loop
space. The corresponding loop space equivariant exterior derivative
is then of the form (2.37),
$$
d_{S} ~=~ d+ i_{S} ~=~  c^{a} \partial_{a} +
{\cal X}_{S}^{a} i_{a} ~=~ d + i_{\dot z} + i_{H}
\eqno (3.4)
$$
where $i_{a}$ again denotes a basis for loop space interior
multiplication which is dual to the $c^{a}$ as in (2.39),
$$
i_{a}(t)c^{b}(t')=\delta_{a}^{b}(t-t')
\eqno (3.5)
$$

In order to evaluate (3.2)
using the supersymmetry determined by (3.4),  we introduce the
following generalization of (3.2),
$$
Z_{\xi} ~=~ \int [dz^{a}] [dc^{a}]
exp\{ i \int\limits_{0}^{T} \vartheta_{a}
\dot z^{a} - H + \frac{1}{2}c^{a}\omega_{ab}c^{b}
+ d_{S} \xi \} \eqno (3.6)
$$
Here $\xi$ is an arbitrary one-form on the loop space.
Using the supersymmetry determined by (3.4), we find [\bk] that if
we introduce a "small" variation
$$
\xi ~\to~ \xi + \delta \xi
\eqno (3.7)
$$
where $\delta \xi$ is a homotopically trivial element in the
subspace
$$
{\cal L}_{S} \delta \xi ~=~ 0
\eqno (3.8)
$$
the path integral (3.6) is invariant under this variation
$$
Z_{\xi} ~=~ Z_{\xi + \delta\xi}
\eqno (3.9)
$$
In particular, if $\xi$ itself is a homotopically trivial
element in the subspace
$$
{\cal L}_{S} \xi ~=~ 0
\eqno (3.10)
$$
we conclude [\bk] that the path integral (3.6) is independent of
$\xi$, and coincides with the original path integral (2.1). The
idea is then to try to evaluate (3.2), (3.6) by selecting
$\xi$ in (3.6) properly, so that the path integral simplifies to
the extent that it can be evaluated exactly.

Since the hamiltonian $H$ in (3.2), (3.6) generates a global
action of $S^{1}$, we conclude that the phase space $\Gamma$
admits a Riemannian structure with a metric
tensor $g_{ab}$ which is Lie-derived\footnotemark\footnotetext{
Locally such a metric tensor always exist in regions where $H$
does not have any critical points. For this, it is sufficient to
introduce local Darboux  coordinates so that the hamiltonian
coincides with one of the coordinates, say $H \sim p_{1}$.  For
$g_{ab}$ we can the select {\it e.g.} $g_{ab} \sim \delta_{ab}$.
However, since we require that (3.11) is valid {\it globally} on
$\Gamma$, for a compact phase space this is equivalent to the
requirement, that $H$ generates the global action of $S^{1}$.}
by the hamiltonian vector field ${\cal
X}_{H}^{a}$,
$$
{\cal L}_{H} g ~=~ 0      \eqno (3.11)
$$
 or in component form,
$$
\partial_{a} {\cal X}_{H} {c} g_{cb} + \partial_{b} {\cal
X}_{H}^{c}g_{ca} + {\cal X}_{H}^{c} \partial_{c} g_{ab} ~=~0
\eqno (3.12)
$$

We can construct such a metric tensor
from an {\it arbitrary} metric tensor on $\Gamma$, by averaging it
over the circle $S^{1} \sim U(1)$. Obviously, this metric tensor
is not unique: For example, if $g_{ab}$ satisfies the
condition (3.11), the following one-parameter generalization of
$g_{ab}$ also satisfies (3.11)
$$
g_{ab} ~\to ~ g_{ab} + \mu \cdot g_{ac}
{\cal X}_{H}^{c} {\cal X}_{H}^{d} g_{db}
\eqno (3.13)
$$

If we select $g_{ab}$ so that it satisfies
(3.11), the following one-parameter family of
functionals is in the subspace (3.10),
$$
\xi_{\lambda} ~=~ \frac{\lambda}{2} g_{ab}{\cal X}_{S}^{a}
c^{b}
\eqno (3.14)
$$
 If a variation of the parameter $\lambda$ indeed determines a
homotopically trivial variation (3.9), the corresponding path integral
(3.6) is independent of $\lambda$ and for $\lambda \to 0$ it reduces to
the original path integral (3.2).  Consequently (3.6), (3.14) coincides
with (3.2) for all values of $\lambda$, and if we evaluate (3.6), (3.14)
in the $\lambda\to\infty$ limit we get the path integral version (2.46)
of the Duistermaat-Heckman integration formula [\bk].

Instead of (3.14), we shall here consider the following [\nt]
one-parameter family of functionals in the subspace (3.10)
$$
\xi_{\lambda} ~=~ \frac{\lambda}{2} g_{ab}
\dot z^{a} c^{b}
\eqno (3.15)
$$
The corresponding path integral (3.6)
is (formally) independent of $\lambda$, and as
$\lambda\to 0$ it reduces to the original path integral
(3.2). We shall now evaluate (3.6), (3.15) in the limit
$\lambda \to \infty$:

Explicitly, the action in (3.6) now reads
$$
S ~=~ \int \{ \frac{\lambda}{2} g_{ab} \dot z^{a}
\dot z^{b}
+ (\vartheta_{a} - \frac{\lambda}{2} g_{ab}{\cal X}_{H}^{b})
\dot z^{a} - H
+ \frac{\lambda}{2} c^{a} ( g_{ab} \partial_{t} +
\dot z^{c}g_{bd}\Gamma^{d}_{ac} ) c^{b}
+ \frac{1}{2} c^{a} \omega_{ab} c^{b}\}
\eqno (3.16)
$$
where $\Gamma^{d}_{ac}$ is the Christoffel symbol
for the metric tensor $g_{ab}$. In order to take the
$\lambda \to \infty$ limit, we expand the variables
$z^{a}(t)$ and $c^{a}(t)$ in some complete set of
states $z_{n}^{a}(t), ~ c_{n}^{a}(t)$ ({\it e.g.} a Fourier
decomposition),
$$
z^{a}(t) ~=~ z^{a}_{o} + \sum\limits_{n\not=0}
s^{a}_{n} z_{n}^{a}(t) ~\sim~ z^{a}_{o} + z^{a}_{t}
\eqno (3.17.a)
$$
$$
c^{a}(t) ~=~ c^{a}_{o} + \sum\limits_{n\not=0}
\sigma^{a}_{n} c^{a}_{n}(t) ~\sim~ c^{a}_{o} + c^{a}_{t}
\eqno (3.17.b)
$$

Here $z^{a}_{o}$ and $c^{a}_{o}$ are the
constant modes of $z^{a}(t)$ and
$c^{a}(t)$ in this expansion, and $z^{a}_{t}$ and $c^{a}_{t}$
denote the remaining $t$-dependent fluctuation modes in the
expansion.  We define the path integral measure in the usual
fashion,
$$
[d z^{a}][dc^{a}] ~=~
dz^{a}_{o} dc^{a}_{o}\prod\limits_{n\not=0} ds^{a}_{n}
\prod\limits_{n\not=0} d\sigma^{a}_{n}
{}~\sim~ d z^{a}_{o} dc^{a}_{o}
\prod\limits_{t} dz^{a}_{t}
dc^{a}_{t}
\eqno (3.18)
$$
We then introduce the change of variables
$$
z^{a}_{t} ~\rightarrow~
\frac{1}{\sqrt{\lambda}} z^{a}_{t}
\eqno (3.19.a)
$$
$$
c^{a}_{t} ~\rightarrow~ \frac{1}{\sqrt{\lambda}} c^{a}_{t}
\eqno (3.19.b)
$$
Formally, the Jacobian for this change of
variables is trivial since the bosonic Jacobian is cancelled by
the fermionic Jacobian. In the $\lambda\to\infty$
limit the path integral over the
fluctuation modes $z_{t}^{a}$ and $c_{t}^{a}$ can be evaluated
explicitly, and the result is
$$
Z ~=~ \int dz^{a}_{o}
dc^{a}_{o} { e^{ -iT H + i\frac{T}{2}
c^{a}_{o} \omega_{ab} c^{b}_{o} }  \over \sqrt{
Det [ \delta_{a}^{b} \partial_{t}  + g^{ac}
( \Omega_{cb} + R_{cb}
) ] }}
\eqno (3.20)
$$
where we have defined
$$
\Omega_{ab} ~=~ \frac{1}{2}\left[
\partial_{b} (g_{ac}{\cal X}^{c}_{H}) -
\partial_{a} (g_{bc} {\cal X}^{c}_{H}) \right]
\eqno (3.21)
$$
which can be identified as the Riemannian momentum map [\bg]
corresponding to the action of ${\cal X}_{H}$ on the Riemannian
manifold $(\Gamma , g_{ab})$  and
$$
R_{ab} ~=~ \frac{1}{2} R_{abcd}c^{c}_{o} c^{d}_{o}
\eqno (3.22)
$$
is the curvature two-form on the phase space $\Gamma$. Here both
(3.21) and (3.22) are evaluated at the constant mode $z_{o}^{a}$.
We evaluate the determinant in (3.20) {\it e.g.}
by the $\zeta$-function method, which yields our final
integration formula
$$
Z ~=~ \int  dz^{a}_{o} dc^{a}_{o}
{}~ e^{ -iT H + i\frac{T}{2}
c^{a}_{o} \omega_{ab} c^{b}_{o} } \sqrt
{Det\left[ {  \frac{T}{2}({\Omega^{a}}_{b} +
{R^{a}}_{b}) \over sinh[ \frac{T}{2}
({\Omega^{a}}_{b} + {R^{a}}_{b})] } \right] }
\eqno (3.23)
$$

We argue, that the integrands in (3.23) determine
{\it equivariant} characteristic classes [\bg] associated with
the equivariant exterior derivative
$$
d_{H} ~=~ d + i_{H} ~=~ c_{o}^{a} {\partial \over \partial
z_{o}^{a} } ~ + ~ {\cal X}_{H}^{a} i_{a}
\eqno (3.24)
$$
which operates on the exterior algebra of the original phase space
$\Gamma$. Indeed, if we set $H=0$ in (3.23), we can identify the
exponential term in (3.23) as the Chern character for the
symplectic two-form $\omega_{o} = c^{a}_{o}\omega_{ab}c^{b}_{o}$,
and the second term in (3.23) yields the $\hat A$-genus for the
curvature two-form $R_{abcd}c^{c}_{o}c^{d}_{o}$.  For $H\not=0$,
the corresponding expressions then determine the equivariant
generalizations of these
characteristic classes:

For the exponential term this can be shown immediately. Since
$$
(d + i_{H}) (H - \frac{1}{2}
c^{a} \omega_{ab}
c^{b}) ~=~ 0
\eqno (3.25)
$$
we conclude that the exponential in (3.23) is equivariantly
closed, hence it determines an equivariant generalization of the
Chern character.

In order to demonstrate that the determinant in (3.23) is an equivariant
generalization of the $\hat A$-genus, we introduce a covariant
generalization of the equivariant exterior derivative,
 $$
D_{H} ~=~ d +
\Gamma + i_{H} ~=~ D + i_{H} \eqno (3.26)
$$
where $\Gamma$ is the
Christoffel symbol one-form.  We then find from the relation
$$
[D_{a},D_{b}]{\cal X}_{c} ~=~ {R_{c}^{d}}_{ab} {\cal X}_{d} +
\Gamma^{d}_{bc} \partial_{a} {\cal X}_{d} - \Gamma^{d}_{ac} \partial_{b}
{\cal X}_{d} \eqno (3.27)
$$
 that the argument in the determinant in
(3.23) is equivariantly covariantly closed,
$$
 D_{H} (\Omega_{ab} +
R_{ab}) ~=~ 0 \eqno (3.28)
$$
 Invoking an equivariant generalization of
an argument which is originally due to Chern [\eg] we then conclude that
the determinant in (3.23) is equivariantly closed,
 $$
(d + i_{H}) \sqrt {Det\left[ { \frac{T}{2}(
{\Omega^{a}}_{b} + {R^{a}}_{b}) \over
sinh[ \frac{T}{2} ({\Omega^{a}}_{b} + {R^{a}}_{b})]
} \right] } ~=~ 0
\eqno (3.29)
$$
Hence it determines an equivariant generalization of
the $\hat A$-genus, and the integration formula (3.23) can be
written as
$$
Z ~=~ \int  dz^{a}_{o} dc^{a}_{o}
{}~ e^{ -iT H + i\frac{T}{2}
c^{a}_{o} \omega_{ab} c^{b}_{o}} {\hat A}\left[ \frac{T}{2}
({\Omega^{a}}_{b} + {R^{a}}_{b}) \right]
$$
$$
= ~ \int Ch\left[ \frac{T}{2}(H - \omega)\right] {\hat A}\left[
\frac{T}{2} ({\Omega^{a}}_{b} + {R^{a}}_{b}) \right]
\eqno (3.30)
$$

Finally, we observe that
the following one-parameter family of functionals also satisfies
the  condition (3.10),
$$
\xi_{\lambda} ~=~ \frac{\lambda}{2} g_{ab} {\cal
X}_{H}^{a} c^{b}  \eqno (3.31)
$$
where
$$
{\cal X}_{H}^{a} ~=~ - \omega^{ab} \partial_{b} H
\eqno (3.32)
$$
 is the hamiltonian vector field determined by $H$, lifted to the loop
space.  Provided the variation of $\lambda$ is indeed a homotopy
transformation that leaves the path integral (3.6) invariant, we then
conclude that (3.6) with (3.31), and (3.2) coincide.  The corresponding
action in (3.6) with (3.31) is
$$
S ~=~ \int \vartheta_{a}{\dot z}^{a} - H - \frac{1}{2}c^{a}
\omega_{ab} c^{a} + \frac{\lambda}{2} g_{ab} \dot z^{a} {\cal
X}_{H}^{b} - \frac{\lambda}{2} g_{ab} {\cal X}_{H}^{a} {\cal
X}_{H}^{b}
+ \frac{\lambda}{2} c^{a}\partial_{a}(g_{bc} {\cal X}_{H}^{b})c^{c}
\eqno (3.33)
$$
If we assume that the critical point set
of $H$ {\it i.e.} zeroes of the hamiltonian vector field (3.3) is
nondegenerate,  we can evaluate the corresponding path integral
in the $\lambda\to\infty$ limit. For this, we again
introduce the expansion (3.17) and the change of variables (3.19),
and we find that for large values of $\lambda$  the path integral
yields
$$
Z ~=~ \int dz_{o}^{a} dc_{o}^{a} exp \{ -iTH + \frac{T}{2}
c_{o}^{a} \omega_{ab} c_{o}^{b} + \frac{\lambda}{2} c_{o}^{a}
\partial_{a} (g_{bc} {\cal X}_{H}^{c}) c_{o}^{b} -
\frac{\lambda}{2} g_{ab} {\cal X}_{H}^{a} {\cal X}_{H}^{b} \}
$$
$$
\times \int [dz_{t}^{a}][dc_{t}^{a}] exp \{ i\int \frac{1}{2}
z_{t}^{a}
\partial_{a}(g_{bc} {\cal X}_{H}^{c})(\delta_{d}^{b}\partial_{t}
- \partial_{d} {\cal X}_{H}^{b}) z_{t}^{d} + \frac{1}{2} c_{t}^{a}
\partial_{a}(g_{bc}{\cal X}_{H}^{b}) c_{t}^{c} + {\cal
O}(\frac{1}{\lambda}) \}
\eqno (3.34)
$$
In the $\lambda\to\infty$ limit the
integrals over the fluctuation modes $z_{t}^{a}$ and $c_{t}^{a}$
can be evaluated exactly, and the integral over the constant
modes $z_{o}^{a}$ localizes to ${\cal
X}_{H}^{a}=0$. This yields
$$
Z ~=~ \sum\limits_{{\cal X}_{H} =
0} { e^{-iTH} \over \sqrt{ det|| \partial_{a} {\cal X}_{H}^{b} ||
} } \sqrt{ det\left[ { \frac{T}{2} \tilde{\Omega^{a}}_{b} \over
sinh(\frac{T}{2} \tilde{\Omega^{a}}_{b}) } \right] } $$
$$
= ~ \sum\limits_{dH = 0} { \sqrt{ det || \omega_{ab} ||} e^{-iTH}
\over \sqrt{ det|| \partial_{a}\partial_{b} H || } }
{\hat A}(\frac{T}{2} \tilde{\Omega^{a}}_{b})
\eqno (3.35)
$$
where
$$
\tilde{\Omega^{a}}_{b}
{}~=~\frac{1}{2} g^{ae}\left(
g_{ec}\omega^{cd}\partial_{b}\partial_{d}H -
g_{bc}\omega^{cd}\partial_{e}\partial_{d}H \right)
\eqno (3.36)
$$
equals (3.21) when evaluated at the critical points ${\cal
X}_{H}=0$.

The final result (3.35) coincides with that obtained in
[\kn], using an argument based on Weinstein's action invariant
[\aw]. For consistency it should also coincide with (3.23), if
we specify (3.23) to a hamiltonian for which the critical point set
${\cal X}_{H} = 0$ is nondegenerate. In order to derive
(3.35) from (3.23), we use our observation that the integrand in
(3.23) is closed with respect to the equivariant exterior
derivative (3.24). Hence we can apply the invariance (3.9) on the
phase space $\Gamma$, to localize (3.23) further to the critical
points of $H$: From our general arguments we conclude, that
the integral (3.23)  coincides with the more general integral
$$
Z_{\xi} ~=~ \int
dz_{o}^{a} dc_{o}^{a} exp \{ -iTH + \frac{i}{2} T
c_{o}^{a}\omega_{ab}c_{o}^{b} + d_{H} \xi \} {\hat A}[
\frac{T}{2} ({\Omega^{a}}_{b} + {R^{a}}_{b}) ]
\eqno (3.37)
$$
provided $\xi$ is an element in the subspace
$$
{\cal L}_{H} \xi ~=~0
\eqno (3.38)
$$
If we select
$$
\xi ~=~ \frac{\lambda}{2} g_{ab} {\cal X}_{H}^{a} c_{o}^{b}
\eqno (3.39)
$$
which is a one-parameter functional in the invariant subspace
(3.38), our general argument (3.9)  implies that (3.37), (3.39) is
$\lambda$-independent, and in the $\lambda\to\infty$ limit we then
get (3.35). However, if the critical point set of
$H$ is {\it degenerate}, we must use the
more general result (3.23) instead.

Finally, we observe that the matrix (3.21) determines a closed
two-form on the phase space $\Gamma$, hence it also
determines a (pre)symplectic stucture on $\Gamma$. If we introduce
the hamiltonian function
$$
{\cal H} ~=~ g_{ab} {\cal X}_{H}^{a} {\cal X}_{H}^{b}
\eqno (3.40)
$$
we then find that the pairs $(H, \omega_{ab})$ and $({\cal H} ,
\Omega_{ab})$ define a bi-hamiltonian structure in the sense that
if the metric tensor $g_{ab}$ satisfies (3.11), the classical
equations of motion for $H$ and $\cal H$ coincide
$$
\dot z^{a} ~=~ \{ H , z^{a} \}_{\omega} ~=~ \{ {\cal H} ,
z^{a} \}_{\Omega}
\eqno (3.41)
$$
Here $\{ ~ , ~ \}_{\omega}$ denotes Poisson bracket with respect
to the symplectic two-form $\omega_{ab}$, and
$\{ ~ , ~ \}_{\Omega}$ denotes Poisson bracket with respect to the
symplectic two-form $\Omega_{ab}$ {\it i.e.} we have
$$
\omega^{ab} \partial_{b}H ~=~ \Omega^{ab}\partial_{b}{\cal H}
\eqno (3.42)
$$
This is consistent with the classical integrability of the
canonical system $(H,\omega_{ab})$.

\vfill\eject

{\bf 4. Loop Space Circle Action And Equivariant Cohomology}

\vskip 0.8cm

In the previous section we have explained, how the path
integral (2.1) can be evaluated exactly for a hamiltonian $H$
that generates the action of $S^{1}$ on the phase space
$\Gamma$, so that there is a metric tensor $g_{ab}$ which
is Lie-derived by $H$. Our evaluation of (2.1) is based
on  loop space equivariant cohomology determined by the equivariant
exterior derivative (3.4)
$$
d_{S} ~=~ d + i_{\dot z} + i_{H}
\eqno (4.1)
$$
This operator does {\it not} correspond to the model
independent action of $S^{1}$ in the {\it loop} space. As explained
in Section 2, for the model independent $S^{1}$ loop space action
we expect an equivariant exterior derivative which is of the form
$$
d_{S} ~=~ d + i_{\dot z}
\eqno (4.2)
$$
Consequently our evaluation of (2.1) can not be directly
related to the general discussion in Section 2.

In order to relate the evaluation of (2.1) to the general model
independent $S^{1}$ loop space formalism that we have developed in
Section 2 it is necessary to reformulate (2.1) in an appropriate
auxiliary field formalism.  We shall now explain, how this auxiliary
field representation of (2.1) is constructed.  Curiously, we find that
our auxiliary fields turn out to coincide with those introduced in
[\mn], to formulate generic Poincare-supersymmetric theories in terms of
model independent $S^{1}$ loop space equivariant cohomology.

The final integration formula that we shall obtain using the
auxiliary field formalism will of course coincide with (3.23).
Consequently from the point of view of hamiltonians that generate
a $S^{1}$ action on the phase space $\Gamma$, the
evaluation of the path integral (2.1)  using auxiliary fields
only confirms that our auxiliary field representation and the
ensuing evaluation of (2.1) are indeed
correct. However, in the following sections we shall find that our
auxiliary field construction can also be extended for hamiltonians
$H$ that do {\it not} generate the action of $S^{1}$ on the
phase space, but are {\it a priori} arbitrary functionals of an
observable, that generates such a $S^{1}$ action. We can then
apply our  auxiliary field representation to evaluate
exactly path integrals for this class of hamiltonians, even though
for these hamiltonians there does {\it not} exist a globally
defined metric tensor $g_{ab}$ such that (3.11) is satisfied.

In order to construct the appropriate auxiliary field
representation of the path integral (3.2),
we first introduce the following representation
of Dirac $\delta$-function,
$$
\delta(x) ~=~ \frac{1}{\pi}\lim_{\alpha \to \infty}
\sqrt{\alpha} e^{-\alpha x^{2}}
\eqno (4.3)
$$
and write
$$
e^{ - i \int\limits_{0}^{T} H(z) } =~
\int [d\phi] \delta(\phi-1) e^{-i \int\limits_{0}^{T} \phi H(z)}
= ~ \lim_{\alpha\to\infty} (\frac{\sqrt{\alpha}}{\pi})^{N} \int
[d\phi] exp \{ i \int\limits_{0}^{T} - \alpha (\phi - 1)^{2} -
\phi H \}
\eqno (4.4)
$$
Here we have introduced
a discretized definition of the path integral measure using a
time lattice with $N\to\infty$ lattice points so that
$$
[d\phi(t)] ~\sim~ \lim\limits_{N\to\infty}\prod\limits_{i=1}^{N}
d\phi(t_{i})
\eqno (4.5)
$$
We then write the path integral (3.2) as
$$
Z ~=~ \lim_{\alpha\to\infty}(\frac{{\sqrt{\alpha}}}{\pi})^{N}
e^{-iT\alpha} \int [dz^{a}] [dc^{a}] [d\phi] exp \{ i
\int\limits_{0}^{T} \vartheta_{a} {\dot z}^{a}  -
\alpha \phi^{2} - \phi(H-2\alpha) + \frac{1}{2}
c^{a}\omega_{ab}c^{b} \}
\eqno (4.6)
$$
We normalize the path integral by
$$
Z ~\to \sqrt{T}\cdot Z
\eqno (4.7)
$$
and identify
$$
\sqrt{T} ~=~ \sqrt{ det || \partial_{t} || }
\eqno (4.8)
$$
where we have used periodic boundary conditions at $t=0$
and $t=T$ in defining $\partial_{t}$, and we have also excluded the
zero mode.
If we introduce an anticommuting variable
$\eta(t)$ and ignore an irrelevant numerical normalization factor,
we can then write the path integral (4.6) as
$$
Z ~=~ \lim_{\alpha\to\infty}
({\sqrt{\alpha}})^{N}
e^{-iT\alpha} \int [dz^{a}] [dc^{a}] [d\phi] [d\eta] exp \{ i
\int\limits_{0}^{T} \vartheta_{a} {\dot z}^{a} + \eta \dot \eta -
\alpha \phi^{2} - \phi(H-2\alpha) + \frac{1}{2}
c^{a}\omega_{ab}c^{b} \}
\eqno (4.9)
$$
Notice that here we do
{\it not} integrate over the  constant mode $\eta_{o}$ of
$\eta(t)$. This constant mode is absent from (4.9) since it would
correspond to the zero mode of $\partial_{t}$.

We introduce the change of variables
$$
\eta(t) ~\to~ \sqrt{\alpha}\cdot \eta(t)
\eqno (4.10)
$$
This yields
$$
Z ~=~ \lim_{\alpha\to\infty}
\sqrt{\alpha}e^{-iT\alpha} \int [dz^{a}][dc^{a}] [d\phi] [d\eta]
exp\{ i \int \vartheta_{a} {\dot z}^{a} + \alpha \eta \dot\eta -
\alpha \phi^{2} - \phi(H-2\alpha) +
\frac{1}{2}c^{a}\omega_{ab}c^{b} \} \eqno (4.11)
$$
Notice that a single overall factor of $\sqrt{\alpha}$ remains in
the measure. It corresponds to the constant mode of $\eta(t)$
which is absent in (4.11).

We shall now proceed to interpret (4.11) in terms of a
model independent $S^{1}$  loop space equivariant cohomology. For
this it is convenient to realize the exterior derivatives and
interior multiplications canonically: We
introduce the following loop space symplectic structure,
$$
\{ \lambda_{a}(t) ,
z^{b} (t') \} ~=~ \{ {\bar c}_{a} (t) , c^{b} (t') \} ~=~
\delta_{a}^{b}(t-t')
\eqno (4.12.a)
$$
$$
\{ \pi(t), \phi(t') \} ~=~ \{ {\cal P}(t) , \eta(t') \} ~=~
\delta(t-t')
\eqno (4.12.b)
$$
and we shall interpret {\it both} $z^{a}(t)$ and
$\eta(t)$ as coordinates in a superloop space. The
corresponding conjugate variables $\lambda_{a}(t)$ and ${\cal
P}(t)$ are then identified as (functional) derivatives with
respect to these coordinates.  We interpret $c^{a}(t)$ as a
basis for superloop space one-forms corresponding to the
coordinates $z^{a}(t)$, and $\phi(t)$ as a superloop space one-form
corresponding to the coordinate $\eta(t)$, and we identify
$\bar c_{a}(t)$ and $\pi(t)$ as a canonical realization
of the corresponding basis for internal multiplication,
$$
\bar c_{a}(t) ~\sim~ i_{a}(t)
\eqno (4.13.a)
$$
$$
\pi(t) ~\sim~ i_{\phi}(t)
\eqno (4.13.b)
$$
that is
$$
i_{a}(t) c^{b}(t') ~\equiv~ \{ {\bar c}_{a}(t) , c^{b} (t') \} ~=~
\delta_{a}^{b}(t-t')
\eqno (4.14.a)
$$
$$
i_{\phi}(t) \phi(t') ~\equiv~ \{ \pi (t) , \phi (t') \} ~=~
\delta (t-t')
\eqno (4.14.b)
$$

With these notations, the superloop space exterior derivative is
$$
d ~=~ c^{a} {\delta \over
\delta z^{a}} + \phi {\delta \over \delta \eta } ~=~ c^{a}
\lambda_{a} + \phi {\cal P}
\eqno (4.15)
$$
and interior multiplication along a vector field which
determines the global, model independent action of $S^{1}$ in the
superloop space is
$$
i_{ \dot z , \dot \eta} ~=~ {\dot z}^{a} {\bar c}_{a} + \dot\eta
\pi
\eqno (4.16)
$$
The following equivariant exterior derivative
$$
Q ~=~ c^{a} \lambda_{a} + \phi {\cal P} - \dot z^{a} {\bar
c}_{a} - \dot \eta \pi
\eqno (4.17)
$$
then corresponds to the global, model independent action of
$S^{1}$, with the Lie derivative
$$
Q^{2} ~=~ {\cal L} ~=~ - {d \over dt}
\eqno (4.18)
$$

Notice that (4.17), (4.18) is a realization of (2.53), (2.54) in
the superloop space  $z^{a}(t), \eta(t)$.

We shall now demonstrate, that the action in the path
integral (4.11) can be formulated in terms of this model
independent loop space $S^{1}$-equivariant cohomology in the
manner we have described in Section 2. For this we introduce
the following canonical transformation in the superloop space,
$$
Q ~\to~
e^{-\Phi} Q e^{\Phi} ~=~ Q + \{ Q , \Phi \} + \frac{1}{2} \{ \{ Q
, \Phi \} , \Phi \} + ...  \eqno (4.19)
$$
Selecting
$$
\Phi ~=~ \eta {\cal
X}^{a}_{H} {\bar c}_{a} ~=~ \eta \omega^{ab} \partial_{b} H \bar
c_{a}
\eqno (4.20)
$$
we then find for the canonically conjugated equivariant
exterior derivative,
$$
Q ~\to~ Q_{S} ~=~ c^{a}\lambda_{a} + \phi {\cal P} + \phi i_{H} -
\eta {\cal L}_{H} - \dot z^{a} {\bar c}_{a} - \dot \eta \pi
\eqno (4.21)
$$
where
$$
i_{H} ~=~ {\cal X}^{a}_{H} {\bar c}_{a}
\eqno (4.22)
$$
is interior multiplication along the hamiltonian vector
field of $H$ lifted to the superloop space, and
$$
{\cal L}_{H} ~=~ d i_{H} + i_{H} d ~=~ {\cal X}^{a}_{H}
\lambda_{a} + c^{a} \partial_{a} {\cal X}_{H}^{b} {\bar c}_{b}
\eqno (4.23)
$$
is the Lie derivative along this superloop space hamiltonian vector
field.

We shall assume, that we have selected the canonical basis
{\it i.e.} the function $\psi$ in (2.8) so that
$$
{\cal L}_{H} \vartheta ~=~0
\eqno (4.24)
$$
As a consequence,
$$
{\cal X}_{H}^{a} \vartheta_{a} ~=~ H + h
\eqno (4.25)
$$
where $h$ is a constant. We then find that the action in (4.11) can
be represented in the form (2.55),
$$
\int \{ \vartheta_{a} {\dot
z}^{a} + \alpha \eta \dot\eta - \alpha \phi^{2} - \phi(H-2\alpha) -
\frac{1}{2}c^{a}\omega_{ab}c^{b}\}
{}~ = ~  \{ Q_{S} , -\vartheta_{a}c^{a} - \alpha \phi \eta
- (2\alpha+h) \eta \}
\eqno (4.26)
$$
which establishes, that in the present
auxiliary field formalism the equivariant cohomology which is
relevant for the path integral (4.11), indeed corresponds to the
model independent action of $S^{1}$ on the superloop space, hence
the path integral (4.11) can be interpreted entirely in the
general framework of the formalism that we have developed
in Section 2.

We shall now evaluate the path integral (4.11). For this, we
introduce the following two-parameter functional in the superloop
space,
$$
\xi(\beta,\gamma) ~=~ \frac{\beta}{2} g_{ab} {\dot z}^{a} c^{b} +
\frac{\gamma}{2} \eta_{t}\phi_{t}
\eqno (4.27)
$$
Here $\beta$ and $\gamma$ are parameters, and the notation
$\eta_{t}$, $\phi_{t}$ means that we have excluded the constant
modes of $\eta(t)$ and $\phi(t)$ in an expansion
$$
\eta(t) ~=~ \eta_{o} + \eta_{t}
\eqno (4.28.a)
$$
$$
\phi (t) ~=~ \phi_{o} + \eta_{t}
\eqno (4.28.b)
$$
with respect to some complete set of functions as in (3.17).
The motivation for excluding $\eta_{o}$ and $\phi_{o}$ in (4.27) is,
that the constant mode $\eta_{o}$ is also absent
in the path integral (4.11).

The functional (4.27) is in the subspace
$$
{\cal L}_{S} \xi ~=~ 0
\eqno (4.29)
$$
Consequently we expect, that if we add the following
term to the action
$$
S ~\to~ S ~+~
\{ Q_{S} ~,~ \frac{\beta}{2} g_{ab} {\dot z}^{a} c^{b} +
\frac{\gamma}{2} \eta_{t}\phi_{t} \}
\eqno (4.30)
$$
the path integral (3.6) corresponding to (4.11) is independent of
the parameters\footnotemark\footnotetext{Notice
that we can not scale the functionals that appear in (4.26) by
overall multiplicative factors, without changing the value of the
path integral. The reason for this is, that  even though these
functionals are also in the subspace (4.29), their variations by
overall multiplicative parameters would not correspond to
"small", homotopically trivial variations [\kn,\mn].} $\beta$ and
$\gamma$. Explicitly, the action in (3.6), (4.2) is
$$
S ~=~ \{ Q_{S} ~,~ - \vartheta_{a} c^{a} - \alpha\eta(\phi+1)
+ \frac{\beta}{2} g_{ab}{\dot z}^{a} c^{b} + \frac{\gamma}{2}
\phi_{t}\eta_{t} \}
$$
$$
=~ \int \{ - \frac{\beta}{2} g_{ab}{\dot z}^{a} {\dot
z}^{b} + \vartheta_{a} {\dot z}^{a} + (\alpha + \frac{\gamma}{2})
\eta \dot\eta + \frac{\beta}{2} \phi g_{ab} {\cal X}_{H}^{a} {\dot
z}^{b} - \alpha \phi^{2} + \frac{\gamma}{2} \phi_{t}^{2}
$$
$$
- \phi (H-2\alpha) + \frac{1}{2} c^{a}\omega_{ab}c^{b} +
\frac{\beta}{2} c^{a}( g_{ab} \partial_{t} + {\dot z}^{c} g_{bd}
\Gamma_{ac}^{d} )c^{b} \}
\eqno (4.31)
$$

In order to evaluate the path integral we
again introduce the expansions (3.17) and (4.28), and the
appropriate definition (3.18) of the path integral measure. We then
change variables according to
$$
z^{a}(t) ~\to~ z_{o}^{a} ~+~ \frac{1}{\sqrt{\beta}}z_{t}^{a}
\eqno (4.32.a)
$$
$$
c^{a}(t) ~\to~ c_{o}^{a} ~+~\frac{1}{\sqrt{\beta}}c_{t}^{a}
\eqno (4.32.b)
$$
$$
\phi(t) ~\to~ \phi_{0} ~+~\frac{1}{\sqrt{\gamma}}\phi_{t}
\eqno (4.32.c)
$$
$$
\eta(t) ~\to~ \eta_{0} ~+~\frac{1}{\sqrt{\gamma}}\eta_{t}
\eqno (4.32.d)
$$
The Jacobians for these changes of variables cancel
each other. Since the path integral is independent of
$\beta$ and $\gamma$, we can evaluate it in the  $\beta, \gamma
\to \infty$ limit.

For large values of $\beta$ and $\gamma$ the action becomes
$$
\int \{ - \frac{1}{2} g_{ab} {\dot z}_{t}^{a} {\dot
z}_{t}^{b} + (\frac{\alpha}{\gamma} + \frac{1}{2}) \eta_{t}
\dot\eta_{t} + \frac{\sqrt{\beta}}{2} ( \phi_{0} +
\frac{1}{\sqrt{\gamma}} \phi_{t}) (g_{ab} {\cal X}_{H}^{b} +
\frac{1}{\sqrt{\beta}} z_{t}^{c} \partial_{c} [g_{ab} {\cal
X}_{H}^{b}]){\dot z}_{t}^{a} - \alpha \phi_{o}^{2} +
(\frac{1}{2} -  \frac{\alpha}{\gamma})\phi_{t}^{2}
$$
$$
- \phi_{o} (H-2\alpha) + \frac{1}{2} c_{o}^{a} \omega_{ab}
c_{o}^{b} + \frac{1}{2} c_{t}^{a}g_{ab}\partial_{t} c_{t}^{b} +
\frac{1}{2} {\dot z}_{t}^{a} z_{t}^{b} R_{ab} ~+~ {\cal
O}(\frac{1}{\sqrt{\beta}} , \frac{1}{\sqrt{\gamma}} )\}
\eqno (4.33)
$$
where $g_{ab}$, ${\cal X}_{H}^{a}$, $H$, $\omega_{ab}$ and
$R_{ab}=\frac{1}{2}R_{abcd}c_{o}^{c} c_{o}^{d}$ are all evaluated
at the constant modes.

According to our general arguments, the path integral must be
independent of $\beta$ and $\gamma$, at least for regular values of
these parameters. We observe that the $\beta,\gamma \to \infty$ limit
is
ambiguous, hence we can {\it not} directly proceed to this
limit: If we first set
$\beta \to\infty$ followed by $\gamma\to\infty$, the final result
is not properly defined. On the other hand, if we set
$\beta,\gamma \to\infty$ while keeping
$\sqrt{\beta}\cdot \gamma^{-1}$ fixed, the final integral is too
complicated to be evaluated in a closed form. A
calculable and well defined limit is obtained if we {\it first} set
$\gamma \to \infty$ and {\it then} $\beta\to\infty$. In this limit
the action (4.33) becomes
$$
\int \{ - \frac{1}{2} g_{ab} {\dot z}_{t}^{a} {\dot
z}_{t}^{b} + \frac{1}{2} \eta_{t}\dot\eta_{t} + \frac{1}{2}
\phi_{o} z_{t}^{a} \partial_{a}[g_{bc} {\cal X}_{H}^{c}] {\dot
z}_{t}^{b} - \alpha\phi_{o}^{2}
$$
$$
- \phi_{o}(H-2\alpha) + \frac{1}{2} \phi_{t}^{2} + \frac{1}{2}
c_{o}^{a}\omega_{ab}c_{o}^{b} + \frac{1}{2} c_{t}^{a} g_{ab}
\partial_{t} c_{t}^{b} + \frac{1}{2} {\dot z}_{t}^{a} z_{t}^{b}
R_{ab} \} \eqno (4.34)
$$
We can now integrate over $z_{t}^{a}$, $c_{t}^{a}$, $\phi_{t}$
and $\eta_{t}$. This yields
$$
Z ~=~ \sqrt{\alpha} \int d\phi_{o}
dz_{o}^{a}dc_{o}^{a} exp\{ - i  T( \alpha \phi_{o}^{2} + \phi_{o}[
H-2\alpha] + \alpha) + \frac{i}{2} T
c_{o}^{a}\omega_{ab}c_{o}^{b}\} \times {\hat A} [
\frac{T}{2}(\phi_{o} {\Omega^{a}}_{b} + {R^{a}}_{b})]
\eqno (4.35)
$$
If we now take the $\alpha\to\infty$ limit and use (4.3), we get
our integration formula (3.24) for the path integral (4.11),
$$
Z ~=~ \int dz_{o}^{a}dc_{o}^{a}
e^{ - i T H - \frac{i}{2} T c_{o}^{a}\omega_{ab}c_{o}^{b} }
{\hat A} [ \frac{T}{2}({\Omega^{a}}_{b} + {R^{a}}_{b})]
\eqno (4.36)
$$

The previous evaluation
establishes that if the hamiltonian $H$ generates the action
of $S^{1}$ in the phase space $\Gamma$, the
corresponding path integral (2.1) can be related to the general,
model independent $S^{1}$-formalism that we have developed in
Section 2, provided we represent it in terms of auxiliary fields
in a superloop space with both commuting and anticommuting
coordinates. We note, that curiously our superloop space and the
corresponding model independent $S^{1}$ equivariant
cohomology coincides with that introduced in [\mn] for generic
supersymmetric theories.

We shall
now proceed to  generalize our superloop space construction to
exact evaluation of the path integral (2.1)  for hamiltonians $H$
that do {\it not} generate the action of $S^{1}$ on the phase space
$\Gamma$, but are in principle {\it arbitrary} functions of an
observable that does generate the action of $S^{1}$. We shall
find, that the final path integrals over the fluctuation modes are
natural generalizations of those encountered here, in
particular these path integrals also admit an auxiliary field
formalism that  corresponds to our model independent loop space
$S^{1}$-equivariant cohomology.  Since we know that in the present
case our approach gives the correct result, we have full
confidence that also in the general case we get
the correct result.

\vfill\eject

{\bf 5. Hamiltonians Quadratic in H}

\vskip 0.8cm

We shall now proceed to evaluate the
path integral (2.1) for a more general class of
hamiltonians. We shall first consider a hamiltonian which is a
quadratic function of an observable $H$, that generates the action
of $S^{1}$ on the phase space $\Gamma$. In our auxiliary field
formalism the pertinent superloop space equivariant cohomology
correponds again to a model independent $S^{1}$ action, and the
path integral can be evaluated exactly using the method that we
developed in the previous section.

A generalization of the Duistermaat-Heckman integration formula
for Lagrangian path integrals where the loop space functional
$S_{B}$ in (2.33) is quadratic in generators of a
Lie algebra has been recently presented in [\ew]. Here we
can not apply this integration formula directly since in our
case the symplectic one-form $\vartheta_{a}\dot z^{a}$ also
appears in the action and consequently the {\it entire} action can
not be presented as a quadractic function of a generator
of some Lie algebra.

We wish to evaluate the path integral
$$
Z ~=~ \int [dz^{a}][dc^{a}] exp\{ i \int\limits_{0}^{T}
\vartheta_{a} {\dot z}^{a} - \frac{1}{4} H^{2} +
\frac{1}{2}c^{a}\omega_{ab}c^{b} \}
\eqno (5.1)
$$
Here $H$ is an observable that generates the action of
$S^{1}$ on the phase space $\Gamma$. Notice that as a
consequence there exists a metric tensor $g_{ab}$ that
satisfies (3.11), but no such metric tensor exists for
the hamiltonian $H^{2}$ that appears in (5.1).

We proceed by following the previous section: Modulo a trivial
normalization factor we write (5.1) as
$$
Z ~= ~ \int [d\phi][d\eta][dz^{a}][dc^{a}] exp
\{ i \int\limits_{0}^{T} \vartheta_{a}{\dot z}^{a} + \eta\dot\eta +
\phi^{2} - \phi H + \frac{1}{2}c^{a}\omega_{ab}c^{b} \}
\eqno (5.2)
$$
Here the integral over the auxiliary field $\phi(t)$ yields the
hamiltonian $H^{2}$ in (5.1). As in (4.7),
we have again normalized the path integral by $\sqrt{T}$ and we have
represented this normalization factor by the integral over the
anticommuting variable $\eta(t)$. Notice
that the constant mode $\eta_{o}$ of $\eta(t)$ is absent
since it corresponds to the zero mode of $\partial_{t}$ that
we have excluded.

The path integral (5.2) is very similar to the
path integral (4.11). As a
consequence we expect that we can apply the superloop
space equivariant cohomology that we have developed in the previous
section, to evaluate (5.2).

We again introduce the equivariant exterior derivative (4.17), that
corresponds to the model independent action of $S^{1}$ on the
superloop space with coordinates $z^{a}(t), ~ \eta(t)$,
$$
Q ~=~ c^{a}\lambda_{a} + \phi {\cal P} - {\dot z}^{a} {\bar c}_{a}
- \dot\eta \pi
\eqno (5.3)
$$
and perform the canonical transformation (4.19), which maps (5.3)
into (4.21),
$$
Q ~\to~  Q_{S} ~=~
c^{a}\lambda_{a} + \phi {\cal P} + \phi {\cal X}_{H}^{a} {\bar
c}_{a} - \eta {\cal L}_{H} - {\dot z}^{a} {\bar c}_{a} - \dot\eta
\pi
\eqno (5.4)
$$
We then find that the action in (5.2) can be
represented as
$$
S ~=~ \{ Q_{S} ~, ~- \vartheta_{a} c^{a} + \eta (\phi - h) \}
\eqno (5.5)
$$
where we have again selected $\vartheta_{a}$ so that (4.25) is
valid.

We introduce the two-parameter functional (4.27), and add to
the action the $Q_{S}$ exact term
$$
S ~\to~ S ~+~ \{ Q_{S} ~,~ \frac{\beta}{2} g_{ab} {\dot z}^{a}
c^{b} + \frac{\gamma}{2} \phi_{t}\eta_{t} \}
\eqno (5.6)
$$
Notice that we have again excluded the constant
modes of $\eta(t)$ and $\phi(t)$, and our final action is now quite
similar to the one in (4.30), (4.31):
$$
S ~=~  \{ Q_{S} ~,~ - \vartheta_{a} c^{a} + \eta (\phi - h)
+\frac{\beta}{2} g_{ab} {\dot z}^{a} c^{b} + \frac{\gamma}{2}
\phi_{t}\eta_{t} \}
\eqno (5.7)
$$
$$
=~ \int \{ - \frac{\beta}{2} g_{ab}{\dot z}^{a}{\dot z}^{b} +
\vartheta_{a} {\dot z}^{a} + (\frac{\gamma}{2} + 1)\eta\dot\eta
+ \frac{\beta}{2} \phi g_{ab} {\dot z}^{a} {\cal X}_{H}^{b}
+ \phi^{2}
$$
$$
+ \frac{\gamma}{2} \phi_{t}^{2} - \phi H +
\frac{1}{2} c^{a} \omega_{ab}c^{b} + \frac{\beta}{2} c^{a}
( g_{ab}\partial_{t} + {\dot z}^{c} g_{bd} \Gamma_{ac}^{d}) c^{b}
\}
\eqno (5.8)
$$
As a consequence we expect that the evaluation of the corresponding
path integral  proceeds in a similar manner:

We introduce the expansions (3.17), (4.28)  and change variables
according to (4.32). For large values of $\beta$ and $\gamma$ we
then find
$$
S~\to~ \int \{ -\frac{1}{2} g_{ab}{\dot z}_{t}^{a} {\dot
z}_{t}^{b} + \frac{1}{2} \eta_{t}\dot\eta_{t} +
\frac{\sqrt{\beta}}{2} ( \phi_{o} + \frac{1}{{\sqrt\gamma}}
\phi_{t})(g_{ab}{\cal X}_{H}^{b} + \frac{1}{\sqrt{\beta}}
z_{t}^{c} \partial_{c}[g_{ab}{\cal X}_{H}^{b}]){\dot z}_{t}^{a} +
\phi_{o}^{2} + \frac{1}{2}\phi_{t}^{2}
$$
$$
- \phi_{o}H + \frac{1}{2}c_{o}^{a}\omega_{ab}c_{o}^{b} +
\frac{1}{2} c_{t}^{a} g_{ab}\partial_{t}c_{t}^{b} + \frac{1}{2}
z_{t}^{a}{\dot z}_{t}^{b} R_{ab} + {\cal O}( \frac{1}{\sqrt{\beta}}
, \frac{1}{\sqrt{\gamma}} ) \}
\eqno (5.9)
$$
where $g_{ab}$, ${\cal X}_{H}^{a}$, $H$, $\omega_{ab}$ and
$R_{ab}$ are evaluated at the constant modes.

We observe that for the {\it fluctuation} modes, (4.33) and (5.9)
are practically identical. Consequently the
evaluation of the corresponding path integrals over
$z_{t}^{a}, c_{t}^{a}, \phi_{t}$ and $\eta_{t}$ proceeds in
exactly the same manner. As in (4.34), we first take the $\gamma
\to \infty$ limit followed by $\beta \to \infty$ limit, and obtain
$$
S ~\to~ \int \{ - \frac{1}{2}g_{ab} {\dot z}_{t}^{a}
{\dot z}_{t}^{b} + \frac{1}{2}\eta_{t}\dot\eta_{t} + \frac{1}{2}
\phi_{o} z_{t}^{a} \partial_{a}(g_{bc}{\cal X}_{H}^{c}) {\dot
z}_{t}^{b} + \phi_{o}^{2} + \phi_{t}^{2} - \phi_{o}H
$$
$$
+ \frac{1}{2} c_{o}^{a}\omega_{ab}c_{o}^{b} + \frac{1}{2}c_{t}^{a}
g_{ab}\partial_{t} c_{t}^{b} + \frac{1}{2} z_{t}^{a}{\dot
z}_{t}^{b}R_{ab}\}
\eqno (5.10)
$$
We then integrate over the fluctuation modes $z_{t}^{a}$,
$c_{t}^{a}$, $\phi_{t}$, $\eta_{t}$. The pertinent functional
determinants can again be computed {\it e.g.} by the
$\zeta$-function method, and the final result is
$$
Z ~=~ \int
d\phi_{o} dz_{o}^{a} dc_{o}^{a} exp\{ i T \phi_{o}^{2} - iT H +
\frac{i}{2}T c_{o}^{a} \omega_{ab}c_{o}^{b} \} {\hat A}\left[
\frac{T}{2}( \phi_{o}{\Omega^{a}}_{b} + {R^{a}}_{b}) \right] )
\eqno (5.11)
$$
Here $\Omega_{ab}$ is defined as in (3.21).

The result (5.11) is our integration formula for a hamiltonian
which is a quadractic function $H^{2}$ of an observable that
generates the action of $S^{1}$ on the phase space $\Gamma$. It is
a remarkably simple expression in terms of the equivariant
characteristic classes, essentially an integral transformation of
the integration formula (3.23) for $H$, and reduces to (3.23) if we
restrict to $\phi_{o} = 1$.

\vfill\eject

{\bf 6. Hamiltonians Which Are Generic Functions of H}

\vskip 0.8cm

We shall now proceed to generalize the previous evaluation of the
path
integral (2.1) for a hamiltonian which is an {\it a priori}
arbitrary function $P(H)$ of an observable $H$ that generates the
action of $S^{1}$ on the phase space $\Gamma$,
$$
Z ~=~ \int [dz^{a}][dc^{a}] exp \{ i \int\limits_{0}^{T}
\vartheta_{a}{\dot z}^{a} - P(H) +
\frac{1}{2}c^{a}\omega_{ab}c^{b} \}
\eqno (6.1)
$$

In order to evaluate (6.1), we first consider the
quantity
$$
exp \{ - i \int P(H) \}
\eqno (6.2)
$$
We shall then {\it assume}, that there exists another function
$\phi(\xi)$ so that we can write (6.2) as a Gaussian
path integral transformation of $\phi(\xi)$,
$$
exp \{ - i \int P(H) \} ~=~ \int [d\xi] exp \{ i \int
\frac{1}{2} \xi^{2} - \phi(\xi) H \}
\eqno (6.3)
$$
We conclude that locally such a function $\phi(\xi)$ can always be
constructed, but there might be obstructions to construct $\phi(\xi)$
globally.

We change variables $\xi ~\to~ \phi$ in (6.3). This yields
$$
= ~ \int [d\phi] \prod\limits_{t}\xi'(\phi) exp \{ i \int
\xi^{2}(\phi) - \phi H \}
\eqno (6.4)
$$
This change of variables $\xi \to \phi$ which maps the Gaussian in
$\xi$ to a nonlinear function of $\phi$, is reminiscent of the
Nicolai transformation [\hn] in supersymmetric theories:  A generic
supersymmetric theory can be  characterized by the existence of a
change of variables that maps the bosonic part of the supersymmetric
action into a Gaussian, and the Jacobian for this change of variables
coincides with the determinant obtained by integrating over the
(bilinear) fermionic part of the supersymmetric action. This suggest
that we should try to identify the Jacobian in (6.4) as a fermionic
integral in a supersymmetric fashion. For this, we first observe that
the multiplet $\phi,\eta$ that we have introduced in the previous
sections is insufficient. In addition, we must also introduce a
further anticommuting field $\bar \eta$, and a bosonic auxiliary
field $\bar \phi$. In order to construct the pertinent supersymmetric
representation, we use (4.8) to write (6.4) as
$$
= ~ \frac{1}{T}\int [d\phi] \prod\limits_{t}\xi'(\phi) det ||
\partial_{t} || exp \{ i \int \xi^{2}(\phi) - \phi H \}
\eqno (6.5)
$$
We then introduce
$$
\prod\limits_{t} \xi'(\phi) det || \partial_{t} || ~=~ \xi_{o}'
\sqrt{ det \left( \begin{array}{cc} 0 & i \xi' \partial_{t} \\
i \xi' \partial_{t} & i\partial_{t}  \end{array} \right) }
\eqno (6.6)
$$
Here $\xi_{o}$  denotes the constant mode of $\xi(\phi)$, which is
left out of the determinant due to the zeromode of $\partial_{t}$.
We use the anticommuting variables $\eta$ and $\bar\eta$ to write the
determinant as an integral, which yields for (6.5)
$$
= ~ \frac{1}{T} \int [d\phi] [d\bar\phi] [d\eta] [d\bar\eta] \xi_{o}'
exp \{ i \int - \frac{1}{2} \bar\phi^{2} + \xi \bar\phi - \phi H +
\frac{1}{2} \bar\eta \dot{\bar\eta} + \bar\eta \xi' \dot \eta \}
\eqno (6.7)
$$
Notice that we again exclude the constant modes of the anticommuting
variables, corresponding to the zeromodes of $\partial_{t}$.

We can now write the original path integral (6.1) as
$$
Z ~=~ \frac{1}{T} \int [dz^{a}] [dc^{a}] [d\phi] [d\bar\phi] [d\eta]
[d\bar\eta]  \xi_{o}' \cdot e^{i \{ Q_{S} , \psi \} }
\eqno (6.8)
$$
Here the loop space equivariant exterior derivative $Q_{S}$ is
$$
Q_{S} ~=~ c^{a} \lambda_{a} + \phi {\cal P} + \bar\phi \bar{\cal P} +
\phi {\cal X}^{a}_{H} {\bar c}_{a} - \eta {\cal L}_{H} - {\dot z}^{a}
{\bar c}_{a}
- \dot\eta \pi - \dot{\bar\eta} \bar\pi
\eqno (6.9)
$$
with $\bar{\cal P}$ and $\bar\pi$ the canonical conjugates of
$\bar\eta$ and $\bar\phi$ respectively, and
$$
\psi ~=~ - \vartheta_{a}c^{a} + (\xi(\phi) -  \frac{1}{2} \bar\phi)
\bar\eta
\eqno (6.10)
$$

Again, the equivariant exterior derivative (6.9) is related to the
model independent form (4.17) in the extended phase space by the
conjugation (4.19), (4.20).

In order to localize the path integral (6.8), we introduce
a two-parameter family of functionals which modifies (6.10) into
$$
\psi ~\to~ \psi + \frac{\beta}{2} g_{ab} \dot z^{a} c^{b} +
\frac{\gamma}{2}
(\phi_{t} \eta_{t} + \bar\phi_{t}\bar\eta_{t})
\eqno (6.11)
$$
Here the constant modes of $\eta$ and $\bar\eta$ are again excluded.
If we now introduce the changes of variables (4.32) in addition of
$$
\bar\phi_{t} ~\to~ \frac{1}{\sqrt{\gamma}} \bar\phi_{t}
\eqno (6.12.a)
$$
$$
\bar\eta_{t} ~\to~ \frac{1}{\sqrt{\gamma}} \bar\eta_{t}
\eqno (6.12.b)
$$
we find in the $\gamma\to\infty$,  $\beta\to\infty$
limit the following integration formula for the path integral (6.1),
$$
Z ~=~ \int dz^{a}_{o} dc^{a}_{o} d\phi_{o} \cdot \xi'_{o} \cdot
e^{-iT\phi_{o}H + \frac{i}{2}T \xi_{o}^{2} +  \frac{i}{2} T
c_{o}^{a}\omega_{ab}c_{o}^{b}}
{\hat A} \left[ \frac{T}{2} (\phi_{o}{\Omega^{a}}_{b} + {R^{a}}_{b} )
\right]
\eqno (6.13)
$$
which is our final integration formula for the path integral (6.1):
It is again remarkably simple expression in terms of the equivariant
characteristic classes, the only complication is that in general it
might not be very easy to identify the function $\xi(\phi)$.

\vfill\eject

{\bf 7. An Example}

\vskip 0.8cm

We shall now verify, that our integration
formula (6.13) indeed yields correct results in simple examples.
For this, we consider the quantization of spin, {\it i.e.}
path integral defined on the co-adjoint orbit $S^{2}$ of SU(2), and
hamiltonian which is a function of the Cartan generator $J_{3}$ of
SU(2). We shall apply the integration formula (6.13) both for $H =
J_{3}$ and for $H = J_{3}^{2}$.

We shall first consider the case $H = J_{3}$: In the spin-$j$
representation of SU(2), the canonical realization of $J_{3}$ on the
Riemann sphere $S^{2}$ is
$$
J_{3} ~\sim~ H ~=~ j { 1-z\bar z \over 1+z \bar z}
\eqno (7.1)
$$
and the corresponding symplectic structure is determined
by the two-form
$$
\omega ~=~ \frac{1}{2} \omega_{ab}c^{a}c^{b} ~=~ { 2 i j \over
(1+z \bar z)^{2} } c^{z}  c^{\bar z}
\eqno (7.2)
$$

In order to apply the integration formula (6.13) in the form (3.30),
we
first evaluate the equivariant curvature two-form,
$$
{\Omega^{a}}_{b} +  {R^{a}}_{b} ~=~ {{\cal R}^{a}}_{b} ~=~
\frac{1}{2}g^{ac} \left[ \partial_{b} (g_{cd} {\cal X}_{H}^{d}) -
\partial_{c} ( g_{bd} {\cal X}_{H}^{d} ) \right] + \frac{1}{2}{{
R^{a}}_{b}}_{cd} c^{c}c^{d}
\eqno (7.3)
$$
Its only non-vanishing components are
$$
{{\cal R}^{z}}_{z} ~=~ - {{\cal R}^{\bar z}}_{\bar z} ~=~ \frac{i}{j}
( H - \omega)
\eqno (7.4)
$$
Consequently we get for the ${\hat A}$-genus
$$
{\hat A}({{\cal R}^{a}}_{b}) ~=~ \frac{T}{2j} \cdot { H - \omega
\over
\sin\left[ \frac{T}{2j} (H-\omega) \right] }
\eqno (7.5)
$$
Explicitly,
$$
H - \omega ~=~ j \left( {1 - z\bar z \over 1+z\bar z} - { 2 i \over
(1+z \bar z)^{2} } c^{z} c^{\bar z} \right) ~=~
j \left( { 1 - z \bar z - c \bar c \over 1 + z \bar z + c \bar c }
\right)
\eqno (7.6)
$$
where we have redefined $c^{a} \to \sqrt{ i } c^{a}$. Hence
the integrand in the integration formula (3.30) is a
function of the combination $z\bar z + c \bar c$ only,
$$
Z ~=~ \frac{i}{\pi T} \int dz d\bar z dc d\bar c ~ F(z \bar z +
c \bar c)
\eqno (7.7)
$$
where from (3.23), (7.5), (7.6) the function $F(y)$ is
$$
F(y) ~=~
{ \frac{T}{2} \cdot { 1 - y \over 1 + y }  \over
\sin \left[ \frac{T}{2} \left( { 1-y \over 1+y } \right) \right] }
\cdot exp \{ -ijT \left( {1 -y \over 1+y } \right) \}
\eqno (7.8)
$$
The integral (7.7) can be evaluated using the Parisi-Sourlas
integration formula
$$
\frac{1}{\pi}\int d^{2}x d\theta d\bar\theta F(x^{2} +
\theta\bar\theta
)  ~=~ \int\limits_{0}^{\infty} du { d F(u) \over du }
{}~=~ F(\infty) - F(0)
\eqno (7.9)
$$
and the result is
$$
Z ~=~ { \sin( Tj ) \over \sin ( \frac{1}{2} T ) }
\eqno (7.10)
$$
If we then introduce the Weyl shift\footnotemark\footnotetext{ In the
literature, it appears that the path integral evaluation  of the Weyl
character for simple Lie groups usually differs from the correct
result
by a Weyl shift, {\it i.e.} in order to get the correct result the
highest weight has to be shifted by half the sum over positive
roots. Several different arguments have been presented to eplain,
why this shift must be introduced, and usually it can be traced back
to
the particular fashion how the path integral is regulated. Here we
are
not concerned with this issue, we are interested in a different
aspect of the evaluation of (2.1). We refer to [\kn,\ot] for further
discussion.} $j \to
j+\frac{1}{2}$, we get the Weyl character formula for SU(2),
$$
Z ~=~ { \sin( T[j+ \frac{1}{2}] ) \over \sin ( \frac{1}{2} T ) }
\eqno (7.11)
$$
and consequently the integration formula (3.30), (6.13) yields the
correct result in this case.

We shall now apply (6.13), (5.11) to evaluate the path integral for
the
hamiltonian $J_{3}^{2}$ in the spin-$j$ representation of SU(2):
$$
H^{2} ~=~  j^{2} \left({ 1-z\bar z \over 1+z \bar z}\right)^{2}
\eqno (7.12)
$$
{}From (7.5), and using (7.6) we find that the corresponding
integration
formula (6.13), (5.11) can be written in the form
$$
Z ~=~ { i \over \sqrt{ 4 \pi i T} } \int\limits_{-\infty}^{\infty}
d\phi \frac{1}{\phi} \int dz d\bar z  dc d \bar c ~ F_{\phi} (z \bar
z
+ c \bar c )
\eqno (7.13)
$$
where
$$
F_{\phi} (y) ~=~  { \frac{T\phi}{2} \cdot { 1 - y \over 1 + y }
\over sin \left[ \frac{T\phi}{2} \left( { 1-y \over 1+y } \right)
\right]
} \cdot exp \{ \frac{i}{4} T \phi^{2} - ijT\phi \left( {1 -y \over
1+y }
\right) \}  \eqno (7.14)
$$
and we have redefined
$$
c^{a} ~\to~ \sqrt{ \frac{i}{\phi} } c^{a}
\eqno (7.15)
$$
 We can again evaluate this integral using the Parisi-Sourlas
integration formula (7.9), and introducing the Weyl shift $j\to
j+\frac{1}{2}$ we get
 $$
Z ~=~ \sqrt{ {T \over 4 \pi i } }\int\limits_{-\infty}^{\infty}
d\phi~  e^{ \frac{i}{4}T \phi^{2} } \cdot { \sin[ (j+\frac{1}{2})
T\phi ]
\over \sin( \frac{1}{2} T \phi) } ~=~
\sum\limits_{m=-j}^{j} \sqrt{ \frac{T}{4\pi i} }
\int\limits_{-\infty}^{\infty} d\phi ~ e^{-iT\phi m} \cdot
e^{\frac{i}{4} T \phi^{2} }
$$
$$
{}~=~
\sum\limits_{m=-j}^{j} e^{-iTm^{2}} ~=~ Tr\{ e^{-iT H^{2}} \}
\eqno (7.16)
$$
which is again the correct result.

\vfill\eject

{\bf 8. Conclusions }

\vskip 0.8cm

In conclusion, we have identified a general class of
hamiltonians for which the path integral can be evaluated
exactly in the sense, that it reduces into an ordinary
integral over the classical phase space of the theory. These
integration formulas are applicable whenever the hamiltonian
is an {\it a priori} arbitrary function of an observable
with $S^{1}\sim U(1)$ action, {\it i.e.} the
hamiltonian can be viewed as a function of a Cartan generator
for some Lie algebra on the phase space. Generically, such
hamiltonians are encountered in classically integrable
models, where the hamiltonians in an integrable hierachy are
functions of the action variables only. Since our
integration formulas have a definite interpretation in terms
of equivariant characteristic classes, we hope that a proper
generalization of our approach might yield a
geometric characterization of quantum integrability.

We also note that we have derived our integration
formulas using a superloop space construction, which is
essentially identical to the interpretation of Poincare
supersymmetric theories in terms of superloop space
equivariant cohomology [\mn]. It would be
interesting to understand, whether this is simply a
coincidence, or if it suggests some deeper  relation between
quantum integrable bosonic theories and supersymmetric
quantum theories.

Finally, we observe that the integrals that we encountered
in Section 7. could all be evaluated in a very simple manner using
the
Parisi-Sourlas integration formula (7.9). We do not have a
general explanation why the integrands exhibit Parisi-Sourlas
supersymmetry. Obviously, it would be very interesting to
establish this Parisi-Sourlas supersymmetry in the
general case already at the level of (6.13) or (2.1).

\vskip  0.6cm

We thank N. Nekrasov for pointing out some mistakes in a preliminary
version, and for explaining his work on
equivariant cohomology and integration formulas. A.N. also thanks
S. Antonov, L. Faddeev,  A. Mironov, A. Morozov, I. Polubin
and A. Reyman for discussions and comments.

\textheight 9.2in

\vfill\eject

\baselineskip 0.55cm
{\bf References}

\vskip 0.5cm

\begin{enumerate}

\item M.A. Semenov-Tyan-Shanskij, Mat. Ind. Steklov (LOMI) {\bf 37}
(1973) 53; and Izv. Akad. Nauk. {\bf 40} (1976) 562

\item J.J. Duistermaat and G.J. Heckman, Inv. Math. {\bf 69} (1982)
259;  and  {\it ibid} {\bf 72} (1983) 153

\item N. Berline and M. Vergne, Duke Math. Journ. {\bf 50} (1983) 539

\item M.F. Atiyah and R. Bott, Topology {\bf 23} (1984) 1

\item J.-M. Bismut, Comm. Math. Phys. {\bf 98} (1985) 213; and {\it
ibid.} {\bf 103} (1986) 127

\item N. Berline, E. Getzler and M. Vergne, {\it Heat Kernels and
Dirac
Operators} (Springer verlag, Berlin, 1991)

\item  M.F. Atiyah, Asterisque {\bf 131} (1985) 43

\item M. Blau, E. Keski-Vakkuri and A.J. Niemi, Phys. Lett.

{\bf B246} (1990) 92;

\item A.J. Niemi and O. Tirkkonen, Phys. Lett. {\bf B293} (1992) 339

\item H.M. Dykstra, J.D. Lykken and E.J. Raiten, preprint
FERMI-PUB-92/383-T (hep-th 9212126)

\item E. Witten, preprint IASSNS-HEP-92/15 (hep-th 9204083)

\item N. Nekrasov, private communication

\item A. Hietam\"aki, A.Yu. Morozov, A.J. Niemi and K. Palo, Phys.
Lett.
{\bf 263B} (1991) 417; A. Yu. Morozov, A.J. Niemi and K. Palo, Phys.
Lett. {\bf 271} (1991) 365; and Nucl. Phys. {\bf B377} (1992) 295

\item V. Guillemin and S. Sternberg, {\it Symplectic Techniques in
Physics}
(Cambridge University Press, 1984)

\item See for example T. Eguchi, P.B. Gilkey and A.J. Hanson, Phys.
Reports {\bf 66}
(1980) 213

\item E. Keski-Vakkuri, A.J. Niemi, G. Semenoff and O. Tirkkonen,
Phys.
Rev. {\bf D44} (1991) 3899

\item A. Weinstein, Math. Z. {\bf 201} (1989) 75

\item H. Nicolai, Phys. Lett. {\bf B89} (1980) 341; Nucl. Phys. {\bf
B176} (1980) 419

\item O. Tirkkonen, preprint HU-TFT-92-45

\end{enumerate}

\end{document}